\documentclass[twocolumn, twoside]{IEEEtran}
\usepackage{amssymb,epsfig,graphics, graphicx, url,times,mathrsfs,algorithm,algorithmic,amsmath,array,latexsym,fancyhdr,xspace,wrapfig, xcolor,multirow,amsthm,caption,cite,helvet,sidecap,bm,hyperref,url}

\usepackage[normalem]{ulem}

\usepackage[inline]{enumitem}
\usepackage{arydshln}
\usepackage[makeroom]{cancel}
\usepackage{varwidth}

\usepackage{tikz}
\usetikzlibrary{shapes.geometric}
\usetikzlibrary{shapes,snakes,patterns}
\usetikzlibrary{arrows,positioning,automata,calc}
\usetikzlibrary{plotmarks}

\definecolor{green1}{rgb}{0,0.5,0}
\definecolor{magenta}{rgb}{1.0, 0.11, 0.81}
\definecolor{mulberry}{rgb}{0.77, 0.29, 0.55}
\definecolor{xgray}{rgb}{0.9, 0.9, 0.9}

\def \bes{\begin{equation*}}
\def \ees{\end{equation*}}
\def \bas{\begin{align*}}
\def \eas{\end{align*}}
\def \be{\begin{equation}}
\def \ee{\end{equation}}
\def \bbm{\begin{bmatrix}}
\def \ebm{\end{bmatrix}}

\def \rvA{\texttt{A}}
\def \rvB{\texttt{B}}
\def \rvC{\texttt{C}}

\def \F{\mathbb{F}}
\newcommand{\Fdata}[1]{\textcolor{blue}{\mathbf{#1}}}
\newcommand{\ftr}[2]{f_{#1}^{(#2)}}
\newcommand{\gtr}[2]{g_{#1}^{(#2)}}
\newcommand{\htr}[2]{h_{#1}^{(#2)}}

\newcommand{\cA}{{\cal A}}

\newcommand{\cF}{{\cal F}}

\newcommand{\cK}{{\cal K}}

\newcommand{\cR}{{\cal R}}

\newcommand{\cW}{{\cal W}}

\newcommand{\cZ}{{\cal Z}}

\newcommand{\mA}{\mathbf{A}}
\newcommand{\mB}{\mathbf{B}}
\newcommand{\mC}{\mathbf{C}}
\newcommand{\mR}{\mathbf{R}}
\newcommand{\mS}{\mathbf{S}}

\newcommand{\mtA}[2]{\widetilde{\mA}_{#1}^{(#2)}}
\newcommand{\mtB}[2]{\widetilde{\mB}_{#1}^{(#2)}}

\newtheorem{theorem}{Theorem}

\newtheorem{lemma}{Lemma}

\newtheorem{definition}{Definition}

\newtheorem{example}{Example}

\usetikzlibrary{decorations.pathreplacing,calc}
\tikzset{brace/.style={decorate, decoration={brace}},
 brace mirrored/.style={decorate, decoration={brace,mirror}},
}

\newcounter{brace}
\newcommand{\fcA}[3]{\widetilde{\mA}_{#1,#2}^{(#3)}}
\newcommand{\fcB}[3]{\widetilde{\mB}_{#1,#2}^{(#3)}}
\def \interval{\Delta}

\IEEEoverridecommandlockouts

\begin{document}

\title{Rateless Codes for Private Distributed Matrix-Matrix Multiplication}

\author{\IEEEauthorblockN{
Rawad Bitar, Marvin Xhemrishi and Antonia Wachter-Zeh }\\ 
\IEEEauthorblockA{Institute for Communications Engineering, Technical University of Munich, Munich, Germany\\
\{\texttt{rawad.bitar, marvin.xhemrishi, antonia.wachter-zeh}\}\texttt{@tum.de}}
\thanks{
This work was partly supported by the Technical University of Munich - Institute for Advanced Studies, funded by the German Excellence Initiative
and European Union Seventh Framework Programme under Grant Agreement
No. 291763.
}
}


\maketitle
\begin{abstract}
We consider the problem of designing rateless coded private distributed matrix-matrix multiplication. A master server owns two private matrices $\mA$ and $\mB$ and wants to hire worker nodes to help compute the multiplication. The matrices should remain private from the workers, in an information-theoretic sense. This problem has been considered in the literature and codes with a predesigned threshold are constructed. More precisely, the master assigns tasks to the workers and waits for a predetermined number of workers to finish their assigned tasks. The size of the tasks assigned to the workers depends on the designed threshold. 

We are interested in settings where the size of the task must be small and independent of the designed threshold. We design a rateless private matrix-matrix multiplications scheme, called RPM3. Our scheme fixes the size of the tasks and allows the master to send multiple tasks to the workers. The master keeps receiving results until it can decode the multiplication. Two main applications require this property: \begin{enumerate*}[label={\textit{\roman*)}}] 
\item leverage the possible heterogeneity in the system and assign more tasks to workers that are faster; and \item assign tasks adaptively to account for a possibly time-varying system.\end{enumerate*}
\end{abstract}

\section{Introduction}

We consider the problem in which a master server owns two private matrices $\mA$ and $\mB$ and wants to compute $\mC=\mA \mB$. The master splits the computation into smaller tasks and distributes them to several worker nodes that can run those computations in parallel. However, waiting for all workers to finish their tasks suffers from the presence of slow processing nodes \cite{DB13,dean2012large}, referred to as \emph{stragglers}, and can outweigh the benefit of parallelism, see e.g., \cite{AGSS13,FASTC,speeding} and references therein. 

Moreover, the master's data must remain private from the workers. We are interested in information-theoretic privacy which does not impose any constraints on the computational power of the compromised workers. On the other hand, information-theoretic privacy assumes that the number of compromised workers is limited by a certain threshold.

We consider applications where the resources of the workers are different, limited and time-varying. Examples of this setting include edge computing in which the devices collecting the data (e.g., sensors, tablets, etc.) cooperate to run the intensive computations. In such applications, the workers have different computation power, battery life and network latency which can change in time. We refer to this setting as heterogeneous and time-varying setting.

We develop a coding scheme that allows the master to offload the computational tasks to the workers while satisfying the following requirements: \begin{enumerate*}[label=\textit{\roman*)}] \item leverage the heterogeneity of the workers, i.e., assign a number of tasks to the workers that is proportional to their resources; \item adapt to the time-varying nature of the workers; and \item maintain the privacy of the master's data.
\end{enumerate*}

We focus on matrix-matrix multiplication since they are a building block of several machine learning algorithms \cite{suykens1999least,seber2012linear}. We use coding-theoretic techniques to encode the tasks sent to the workers. We illustrate the use of codes to distribute the tasks in the following example.

\begin{table*}[t]
    \centering
    \renewcommand{\arraystretch}{1.2}
    \resizebox{\textwidth}{!}{
    \begin{tabular}{c|c|c|c|c|c}
    ~ & Worker~$1$ & Worker~$2$ & Worker~$3$ & Worker~$4$ & Worker~$5$ \\ \hline%
      \multirow{2}{*}{Round $1$} & $\mR_1(1-a_1)+\Fdata{A_1} a_1$ &  $\mR_1(1-a_2)+\Fdata{A_1} a_2$ &    $\mR_1(1-a_3)+\Fdata{A_1} a_3$&  $\mR_1(1-a_4)+\Fdata{A_2} a_4$&  $\mR_1(1-a_5)+\Fdata{A_2} a_5$ \\ %
       ~ & $\mS_1(1-a_1)+\Fdata{B} a_1$ &  $\mS_1(1-a_2)+\Fdata{B} a_2$ &  $\mS_1(1-a_3)+\Fdata{B} a_3$&  $\mS_1(1-a_4)+\Fdata{B} a_4$&   $\mS_1(1-a_5)+\Fdata{B} a_5$ \\[0.5em] \hdashline%
      \multirow{2}{*}{Round $2$} & $\mR_2(1-a_1)+(\Fdata{A_1+A_2}) a_1$ &  $\mR_2(1-a_2)+(\Fdata{A_1+A_2}) a_2$ &    $\mR_2(1-a_3)+(\Fdata{A_1+A_2}) a_3$& ~& ~ \\ %
       ~ & $\mS_2(1-a_1)+\Fdata{B} a_1$ &  $\mS_2(1-a_2)+\Fdata{B} a_2$ &  $\mS_2(1-a_3)+\Fdata{B}u a_3$& ~ &   ~\\
    \end{tabular}
    }
    
    \caption{A depiction of the tasks sent to the workers in Example~\ref{ex:intro2}.}
    \label{tab:intro-ex2}
    \vspace{-0.5cm}
\end{table*}

\begin{example}\label{ex:intro}
Let $\mA \in \F_q^{r\times s}$ and $\mB \in \F_q^{s\times \ell}$ be two private matrices owned by the master who wants to compute $\mC=\mA \mB$. The master has access to $5$ workers. At most $2$ workers can be stragglers. The workers do not collude, i.e., the workers do not share with each other the tasks sent to them by the master. 
To encode the tasks, the master generates two random matrices $\mR \in \F_q^{r\times s}$ and $\mS \in \F_q^{s\times \ell}$ uniformly at random and independently from $\mA$ and $\mB$. The master creates two polynomials\footnote{The multiplication and addition within the polynomials is element-wise, e.g., each element of $\mA$ is multiplied by $x$.} $f(x) = \mR(1-x)+ \mA x$ and $g(x) = \mS(1-x)+\mB x$. The task sent to worker $i$ is $f(a_i)$, and $g(a_i)$,  $i=1,\dots,5$, where $a_i\in \F_q\setminus\{1\}$. Each worker computes $h(a_i) \triangleq f(a_i) g (a_i) = \mR\mS(1-a_i)^2 + \mR\mB a_i(1-a_i) + \mA\mS a_i(1-a_i) + \mA\mB a_i^2$ and sends the result to the master. When the master receives three evaluations of $h(x) \triangleq f(x)g(x)$, it can decode the whole polynomial of degree $2$. In particular, the master can compute $\mA\mB=h(1)$. The privacy of $\mA$ and $\mB$ is maintained because each matrix is padded by a random matrix before being sent to a worker.
\end{example}

In Example~\ref{ex:intro}, even if there are no stragglers, the master ignores the responses of two workers. In addition, all the workers obtain computational tasks of the same complexity\footnote{Each evaluation of the polynomial $f(x)$ (or $g(x)$) is a matrix of the same dimension as $\mA$ (or $\mB$). The computational complexity of the task is therefore proportional to the dimension of the created polynomial.}.

We highlight in Example~\ref{ex:intro2} the main ideas of our scheme that allow the master to assign tasks of different complexity to the workers and use all the responses of the non stragglers.

\begin{example}\label{ex:intro2}
Consider the same setting as in Example~\ref{ex:intro}. Assume that workers $1,2$ and $3$ are more powerful than the others. The master splits $\mA$ into $\mA= \begin{bmatrix}\mA_1^T & \mA_2^T
\end{bmatrix}^T$ and 
wants $\mC=\begin{bmatrix}(\mA_1\mB)^T & (\mA_2\mB)^T
\end{bmatrix}^T$. 
The master divides the computations into two rounds. In the first round, the master generates two random matrices $\mR_1 \in \F_q^{r/2\times s}$ and $\mS_1 \in \F_q^{s\times \ell}$ uniformly at random and independently from $\mA$ and $\mB$. The master creates four polynomials: 
\begin{align*}
  f_1^{(1)}(x) & = \mR_1(1-x)+ \mA_1x,\\
  \ftr{1}{2}(x) & = \mR_1(1-x)+\mA_2 x,\\
  \gtr{1}{1}(x) & =\gtr{1}{2}(x) = \mS_1(1-x)+\mB x.   
\end{align*}
The master sends $\ftr{1}{1}(a_i)$ and $\gtr{1}{1}(a_i)$ to workers $1,2,3$, and sends $\ftr{1}{2}(a_i)$ and $\gtr{1}{2}(a_i)$ to workers $4,5$, where $a_i\in \F_q\setminus\{0,1\}$. Workers $1$, $2$, $3$ compute $\htr{1}{1}(a_i) \triangleq \ftr{1}{1}(a_i) \gtr{1}{1} (a_i)$ and workers $4$, $5$ compute $\htr{1}{2}(a_i)\triangleq \ftr{1}{2}(a_i) \gtr{1}{2} (a_i)$.

The master starts round $2$ when workers $1,2,3$ finish their tasks. It generates two random matrices $\mR_2 \in \F_q^{r/2\times s}$ and $\mS_2 \in \F_q^{s\times \ell}$ and creates $f_2^{(1)}(x) = \mR_2 (1-x) + (\mA_1+\mA_2)x$, $\gtr{2}{1}(x) = \mS_2(1-x)+\mB x$ and sends evaluations to the first three workers which compute $\htr{2}{1}(x)\triangleq \ftr{2}{1}(x)\gtr{2}{1}(x)$. One main component of our scheme is to generate $\widetilde{\mA}_1\triangleq \mA_1$, $\widetilde{\mA}_2\triangleq \mA_2$ and $\widetilde{\mA}_3\triangleq \mA_1+\mA_2$ as Fountain-coded~\cite{Fountain,LT,Raptor} codewords of $\mA_1$ and $\mA_2$.
The tasks sent to the workers are depicted in Table~\ref{tab:intro-ex2}.

\textbf{Decoding $\mC$:} The master has two options: \begin{enumerate*}[label=\textit{\arabic*)}] \item workers $4$ and $5$ finish their first task before workers $1,2,3$ finish their second tasks, i.e., no stragglers. The master interpolates $\htr{1}{1}(x)$ and obtains $\htr{1}{1}(1) = \mA_1\mB$ and $\htr{1}{1}(0) = \mR_1\mS_1$. Notice that  $\htr{1}{2}(0)=\htr{1}{1}(0) = \mR_1\mS_1$. Thus, the master also has three evaluations of $\htr{1}{2}(x)$ and can obtain $\mA_2\mB$. \item workers $4$ and $5$ are stragglers and do not finish their first task before workers $1,2,3$ finish their second tasks. The master interpolates (decodes) both $\htr{1}{1}(x)$ and $\htr{2}{1}(x)$. In particular, the master obtains $\mA_1\mB = \htr{1}{1}(1)$ and $\mA_2\mB = (\mA_1+\mA_2)\mB - \mA_1\mB= \htr{2}{1}(1) - \mA_1\mB.$ \end{enumerate*}
The privacy of $\mA$ and $\mB$ is maintained because each matrix is padded by a different random matrix before being sent to a worker.
\end{example}
%
%
%
{\em Related work:} The use of codes to mitigate stragglers in distributed linear computations was first proposed in \cite{speeding} without privacy constraints. Several works such as \cite{mallick2018rateless,baharav2018straggler,wang2018coded,yu2017polynomial,li2016fundamental,yu2018straggler,fahim2017optimal,KS18,factored_lt} propose different techniques improving on \cite{speeding} and provide fundamental limits on distributed computing. 
Straggler mitigation with privacy constraints is considered in \cite{BPR17,bitar2019private,yang2018secure,d2018gasp,chang2018capacity,kakar2018rate,aliasgari2019distributed,yu2018lagrange,yu2020entangled,kim2019private}. The majority of the literature assumes a threshold of fixed number of stragglers. 
In \cite{BPR17,bitar2019private} the authors consider the setting in which the number of stragglers is not known a priori and design schemes that can cope with this setting. However, \cite{BPR17,bitar2019private} consider the matrix-vector multiplication setting in which only the input matrix must remain private. Our proposed scheme can be seen as a generalization of the coding scheme in \cite{bitar2019private} to handle matrix-matrix multiplication. 

{\em Contributions:} We present a rateless coding scheme for private matrix-matrix multiplication. Our method is based on dividing the input matrices into smaller parts and encode the small parts using rateless Fountain codes. The Fountain-coded matrices are then encoded into small computational tasks (using several Lagrange polynomials) and sent to the workers. The master adaptively sends tasks to the workers. In other words, the master first sends a small task each worker and then starts sending new small tasks to workers who finished their previous task. We show that our scheme satisfies the following properties: \begin{enumerate*}[label={\textit{\roman*)}}]\item it maintains the privacy of the input matrices against a given number of colluding workers; \item it leverages the heterogeneity of the resources at the workers; and \item it adapts to the time-varying resources of the workers\end{enumerate*}. 

\section{Preliminaries}\label{sec:Preliminaries}
We set the notation and define the problem setting. 

\emph{Notation:} For any positive integer $n$ we define $[n]\triangleq \{1,\dots,n\}$. We denote by $n$ the total number of workers. For $i\in[n]$ we denote worker $i$ by $w_i$. 
For a prime power $q$, we denote by $\F_q$ the finite field of size $q$. We denote by $H(\rvA)$ the entropy of the random variable $\rvA$ and the mutual information between two random variables $\rvA$ and $\rvB$ by $I(\rvA;\rvB)$. All logarithms are to the base $q$.

\emph{Problem setting:} The master possesses two private matrices $\mA\in\F_q^{r\times s}$ and $\mB\in \F_q^{s\times \ell}$ uniformly distributed over their respective fields and wants to compute $\mC=\mA\mB \in \F_q^{r \times \ell}$. The master has access to $n$ workers that satisfy the following properties: \begin{enumerate*}[label=\textit{\arabic*)}] \item The workers have different resources. They can be grouped into $c>1$ clusters with $n_u$ workers, $u=1,\dots,c,$ with similar resources such that $\sum_{u\in[c]} n_u = n$. \item The resources available at the workers can change with time. Therefore, the size of the clusters and their number can change throughout the multiplication of $\mA$ and $\mB$.\item The workers have limited computational capacity. \item Up to $z$, $\displaystyle 1\leq z<\min_{u\in[c]}{n_u}$, workers collude to obtain information about $\mA$ and/or $\mB$. If $z = 1$, we say the workers do not collude. 
\end{enumerate*}

The master splits $\mA$ row-wise and $\mB$ column-wise into $m$ and $k$ smaller sub-matrices, respectively, i.e., $\mA=\bbm \mA_1^T, \dots, \mA_m^T\ebm^T$, and $\mB=\bbm \mB_1, \dots, \mB_k\ebm$. The master sends several computational tasks to each of the workers such that each task has the same computational complexity as $\mA_i \mB_j$, $i\in [m], j\in [k]$. After receiving enough responses from the workers, the master should be able to compute $\mC=\mA\mB$.

\begin{definition}[Double-sided $z$-private matrix-matrix multiplication scheme]\label{def:doubly_private}
We say that a matrix-matrix multiplication scheme is double-sided $z$-private if any collection of $z$ colluding workers learns nothing about the input matrices involved in the multiplication. 
Let $\rvA$ and $\rvB$ be the random variables representing the input matrices. We denote by $\cW_i$ the set of random variables representing all the tasks assigned to $w_i$, $i=1,\dots,n$. For a set $\mathcal{A}\subseteq [n]$ we define $\cW_\mathcal{A}$ as the set of random variables representing all tasks sent to workers indexed by $\mathcal{A}$, i.e., $\cW_\mathcal{A}=\{\cW_i| i\in \cA\}$. Then the privacy constraint can be expressed as
\begin{equation}\label{eq:privacy}
    {I}\left(\rvA,\rvB;\cW_\mathcal{Z}\right) = 0, \forall \cZ \subset [n], \text{ s.t. } |\cZ| = z.
\end{equation}

Let $\cR_i$ be the set of random variable representing all the computational results of $w_i$ received at the master. Let $\rvC$ be the random variable representing the matrix $\mC$. The decodability constraint can be expressed as
\begin{equation}\label{eq:decoding}
    {H}\left(\rvC|\cR_1,\ldots,\cR_n\right) = 0.
\end{equation}
Note that the sets $\cR_i$ can be of different cardinality, and some may be even empty, reflecting the heterogeneity of the system and the straggler tolerance.
\end{definition}

Let the \emph{download rate}, $\rho$, of the scheme be defined as the ratio between the number of needed tasks to compute $\mC$ and the number of responses sent by the workers to the master, %

\begin{equation*}
\rho = \frac{mk}{\text{number of received responses}}.
\end{equation*}
We are interested in designing \emph{rateless} double-sided $z$-private codes for this setting. By \emph{rateless}, we mean that the download rate, or simply rate, of the scheme is not fixed a priori, but it changes depending to the resources available at the workers.
For instance, the rate of the scheme in Example~\ref{ex:intro} is fixed to $1/3$, whereas the rate of the scheme in Example~\ref{ex:intro2} is either $2/5$ or $1/3$ depending on the behavior of the workers.




\section{RPM3 Scheme}

We provide a detailed explanation of our RPM3 (Rateless~Private Matrix-Matrix Multiplication) scheme and prove the following theorem.
\begin{theorem}\label{thm:RPM3}
    Consider a matrix-matrix multiplication setting as described in Section~\ref{sec:Preliminaries}. The RPM3 scheme defined next is a rateless double-sided $z$-private matrix-matrix multiplication scheme that adapts to the heterogeneous behavior of the workers.
\end{theorem}

\begin{IEEEproof}
The proof is constructive. We give the details of the construction in Sections~\ref{sec:data_enc} and~\ref{subsec:clustering}. In Section~\ref{sec:decoding} we show that the master can obtain the desired the computation. We prove the privacy constraint in Section~\ref{sec:privacy}. 
\end{IEEEproof}

\subsection{Data encoding}\label{sec:data_enc}
The master divides the encoding into rounds. At a given round $t$, the workers are grouped into $c$ clusters each of $n_u$ workers, $u=1,\dots,c$ and $\sum_{u=1}^c n_u = n$. We shall defer the clustering technique to the next section. We define $d_1 \triangleq \lfloor \frac{n_1 - 2z + 1}{2}\rfloor$ and $d_u \triangleq \lfloor \frac{n_u - z + 1}{2}\rfloor$ for $u = 2,\dots, c$. The master generates $c$ Lagrange polynomial pairs $\ftr{t}{u}(x)$ and $\gtr{t}{u}(x)$. Each polynomial $\ftr{t}{u}(x)$ contains $d_u$ Fountain-coded matrices $\widetilde{\mA}_{t,\kappa}^{(u)}$, $\kappa = 1,\dots, d_u$, defined as\footnote{Note that $b^{(u)}_{\kappa,i}$ also depends on $t$, but we remove the subscript $t$ for the ease of notation.} $\widetilde{\mA}_{t,\kappa}^{(u)} \triangleq \sum_{i=1}^m b_{\kappa,i}^{(u)} \mA_i$, where $b_{\kappa,i}^{(u)} \in \{0,1\}$. Similarly, each polynomial $\gtr{t}{u}(x)$ contains $d_u$ Fountain-coded matrices $\widetilde{\mB}_{t,\kappa}^{(u)} \triangleq \sum_{j=1}^k b_{\kappa,j}^{(u)} \mB_j$ where  $b_{\kappa,j}^{(u)} \in \{0,1\}$ are chosen randomly \cite{LT}. The master generates $2z$ uniformly random matrices $\mR_{t,1},\dots, \mR_{t,z} \in \F_q^{r/m\times s}$ and $\mS_{t,1},\dots, \mS_{t,z} \in \F_q^{s\times \ell/k}$.

Let $d_{\text{max}} = \max_u d_u$ and $\alpha_\delta \in \F_q$ for $\delta \in [d_\text{max}+z]$ be distinct elements of $\F_q$.  The polynomials are constructed as shown in~\eqref{eq:fx} and~\eqref{eq:gx}.
\begin{align}
\ftr{t}{u}(x) &= \sum_{\delta=1}^{z} \mR_{t,\delta} \prod_{\nu \in [{d_u}+z]\setminus \{\delta\}} \frac{x - \alpha_\nu}{\alpha_\delta-\alpha_\nu} \nonumber \\
 &+ \sum_{\delta=z+1}^{{d_u}+z} \widetilde{\mA}^{(u)}_{t,\delta-z} \prod_{\nu \in [d_u+z]\setminus \{\delta\}} \frac{x - \alpha_\nu}{\alpha_\delta-\alpha_\nu}, \label{eq:fx}\\
\gtr{t}{u}(x) &= \sum_{\delta=1}^{z} \mS_{t,\delta} \prod_{\nu \in [d_u+z]\setminus \{\delta\}} \frac{x - \alpha_\nu}{\alpha_\delta-\alpha_\nu} \nonumber \\
&+ \sum_{\delta=z+1}^{d_u+z} \widetilde{\mB}^{(u)}_{t,\delta-z} \prod_{\nu \in [d_u+z]\setminus \{\delta\}} \frac{x - \alpha_\nu}{\alpha_\delta-\alpha_\nu}. \label{eq:gx}
\end{align}

The master chooses $n$ distinct\footnote{Choosing the $\beta_i$'s carefully is needed to maintain the privacy constraints as explained in the sequel.} elements $\beta_i\in \F_q\setminus\{\alpha_1,\cdots,\alpha_{d_{\text{max}+z}}\}$, $i=1,\dots,n$. 
For each worker, $w_i$ the master checks the cluster $u$ to which this worker belongs, and sends $\ftr{t}{u}(\beta_i)$, $\gtr{t}{u}(\beta_i)$ to that worker. 



\subsection{Clustering of the workers and task distribution}\label{subsec:clustering}

\noindent{\em Clustering: }For the first round $t=1$, the master groups all the workers in one cluster of size $n_1 = n$. The master generates tasks as explained above and sends them to the workers. 

For $t>1$, the master wants to put workers that have similar response times in the same cluster. In other words, workers that send their results in round $t-1$ to the master within a pre-specified interval of time will be put in the same cluster. Let $\interval$ be the length of the time interval desired by the master.

In addition to the time constraint, the first cluster must satisfy $n_1 \geq 2z-1$ workers and all other clusters must satisfy $n_u \geq z+1$ workers $u= 2,\dots,c$. Those constraints ensure that the master can decode the respective polynomials $\htr{t}{u}(x)$ as explained in the next section.

Let $\eta_1$ be\footnote{In this section, all variables depend on $t$. However, we omit $t$ for the clarity of presentation.} the time spent until the result of $w_{i_1}$ is received by the master (at round $t-1$). All workers that send their results before time $\eta_1 + \interval$ are put in cluster $1$. If $n_1\geq 2z-1$, the master moves to cluster $2$. Otherwise, the master increases $\interval$ so that $n_1 \geq 2z-1$. The master repeats the same until putting all the workers in different clusters guaranteeing $n_u\geq z+1$, $u=2,\dots,c$.

Over the course of the computation process, the master keeps measuring the empirical response time of the workers. The response time of a worker is the time spent by that worker to receive, compute and return the result of one task. Having those measurements, the master can update the clustering accordingly when needed using the same time intervals.

{\em Task distribution: }
At the beginning of the algorithm, the master generates tasks assuming all workers are in the same cluster and sends those tasks to the workers. For round $2$ the master arranges the workers in their respective clusters and sends tasks accordingly. Afterwards, when the master receives\footnote{To avoid idle time at the workers, the master can measure the expected computation time of each worker at round $t_i-1$. Using this information, the master can then send a task to a worker in a way that this worker will receive the task right after finishing its current computation. This will guarantee that the worker will not be idle during the transmission of tasks to and from the master. See \cite{KS18} for more details.} a task from worker $w_i$, it checks at which round $t_i$ this worker is (how many tasks did the worker finish so far) and to which cluster $u$ it belongs. The  master generates $\ftr{t_i+1}{u}(x), \htr{t_i+1}{u}(x)$ if $w_i$ is the first worker of cluster $u_i$ to finish round $t_i$ and sends $\ftr{t_i+1}{u}(\beta_i), \htr{t_i+1}{u}(\beta_i)$ to $w_i$.

\subsection{Decoding}\label{sec:decoding}
At a given round $t$, the master first waits for the $n_1$ fastest workers belonging cluster $1$ to finish computing their tasks so that it can interpolate $\htr{t}{1}(x)$. This is possible because the master obtains $n_1 = 2d_1+2z-1$ evaluations of $\htr{t}{1}(x)$ equal to the degree of $\htr{t}{1}(x)$ plus one. By construction, for a given $t$, the polynomials $\ftr{t}{u}(x)$ and $\gtr{t}{u}(x)$ share the same random matrices as coefficients, see~\eqref{eq:fx} and~\eqref{eq:gx}. Thus, for $\zeta=1,\dots,z$, the polynomials $\htr{t}{u}(x)$ share the following $z$ evaluations
\begin{equation}
    \htr{t}{1}(\alpha_{\zeta})=\htr{t}{2}(\alpha_{\zeta})=\dots=\htr{t}{c}(\alpha_{\zeta})=R_{t,\zeta}S_{t,\zeta}.
\end{equation}

Therefore, the master can interpolate $\htr{t}{u}(x)$ when $n_u$ workers of cluster $u, u=2,\dots,c,$ return their results. This is possible because the master receives $n_u=2d_u+z-1$ evaluations of $\htr{t}{u}(x)$ and possesses the $z$ evaluations shared with $\htr{t}{1}(x)$. Allowing the polynomials to share the randomness enables us to reduce the number of workers from every cluster $u>1$ by $z$ workers.


After successfully interpolating a polynomial $\htr{t}{u}(x)$ for a given round $t$ and a cluster $u$, the master computes $d_u$ products of Fountain-coded matrices
\begin{equation}
\htr{t}{u}(\alpha_{\kappa+z}) = \mtA{t,\kappa}{u}\mtB{t,\kappa}{u}
\end{equation}
for $\kappa=1,\dots,d_u$. The master feeds those $d_u$ computations to a peeling decoder \cite{LT,factored_lt,Raptor,Fountain} and continues this process until the peeling decoder can successfully decode all the components of the matrix $\mC$.

\subsection{Proof of double-sided privacy}\label{sec:privacy}
Since the master generates new random matrices at each round, it is sufficient to prove that the privacy constraint given in~\eqref{eq:privacy} holds at each round separately. The proof is rather standard and follows the same steps as~\cite{bitar2019private,yu2018lagrange}. We give a complete proof in the Appendix for completeness and provide next a sketch of the proof.

Let $\cW_{i,t}$ be the set of random variables representing the tasks sent to worker $w_i$ at round $t$. For a set $\mathcal{A}\subseteq [n]$ we define $\mathcal{W}_{\mathcal{A},t}$ as the set of random variables representing the tasks sent to the workers indexed by $\mathcal{A}$ at round $t$, i.e., $\mathcal{W}_{\mathcal{A},t}\triangleq \{\cW_{i,t}| i\in \mathcal{A}\}$. We want to prove that at every round $t$ 
\begin{equation}\label{eq:privacy_per_round}
    {I}\left(\rvA,\rvB;\cW_{\mathcal{Z},t}\right) = 0, \forall \cZ \subset [n], \text{ s.t. } |\cZ| = z.
\end{equation}

To prove~\eqref{eq:privacy_per_round} it is enough to show that given the input matrices $A$ and $B$, any collection of $z$ workers $w_{i_1}, \dots,w_{i_z},$ can use the tasks given to them at round $t$ to obtain the random matrices $\mR_{t,1},\dots,\mR_{t,z}$ and $\mS_{t,1},\dots,\mS_{t,z}$.

Proving that a collection of $z$ workers $w_{i_1}, \dots,w_{i_z},$ can obtain the random matrices $\mR_{t,1},\dots,\mR_{t,z}$ and $\mS_{t,1},\dots,\mS_{t,z}$ given their tasks and $\mA$ and $\mB$ follows from the use of Lagrange polynomials.

\section{Rate Analysis}
We analyse the rate of our RPM3 scheme for the special case where the resources of the workers in different clusters are proportional to each other, e.g., the resources of workers in cluster $1$ are three times higher than those of workers in cluster $2$. 
In addition, we assume that the workers in the same clusters have very similar response time. 
We compare RPM3 to the scheme in \cite{kakar2018rate} that has an improved rate over using the Lagrange polynomials but does not exist for all values of $m$ and $k$.

{\em Rate of RPM3: }Let $\tau_u$ be the number of rounds finished (tasks successfully computed) by all the workers in cluster $u$, $u=1,\dots,c$. Under this assumption, there exist integers $\gamma_{u_1,u_2} \geq 1$ for $u_1,u_2 \in [c], u_1< u_2,$ such that $\tau_{u_1} = \gamma_{u_1,u_2} \tau_{u_2}$. This means that the number of tasks computed by workers in cluster $u_1$ is $\gamma_{u_1,u_2}$ times more than the number of tasks computed by workers in the slower cluster $u_2$.

\begin{lemma}\label{lemma:rate}
    Consider a private distributed matrix-matrix multiplication with $n$ workers out of which at most $z$ can collude. Let the input matrices $\mA$ and $\mB$ be split into $m$ and $k$ sub matrices, respectively. 
    
    Let $c$ be the number of clusters of workers and $\tau_u$ be the number of rounds in which the polynomial $\htr{t}{u}(x), t=1,\dots,\tau_u$ is interpolated at the master. Then, for a $\varepsilon$ overhead required by the Fountain code decoding process, the rate of the RPM3 scheme under the special case described above is
    \begin{equation}
        \rho = \dfrac{mk}{2mk(1+\varepsilon)+ (z-1)\tau_c\sum_{u=1}^c\gamma_{u,c} + z\tau_c \gamma_{1,c}}.
    \end{equation}
 \end{lemma}
 We defer the proof of Lemma~\ref{lemma:rate} to the end of this section.
 
 Lemma~\ref{lemma:rate} shows a tradeoff between the rate of the scheme and its adaptivity to heterogeneous systems. Dividing the workers into $c$ clusters and sending several polynomials to the workers affects the rate of the scheme. The loss in the rate appears in the term $(z-1)\tau_c\sum_{u=2}^c\gamma_{u,c}$. However, sending several polynomials to the workers allows the master a flexibility in assigning a number of tasks proportional to the resources of the workers; Hence, increasing the speed of the computing process.
 
 The main property of RPM3 is that the rate of the scheme is independent of the degree of the encoding polynomials and from the number of available workers $n$. The rate only depends on the number of assigned tasks to the workers in different clusters. This property reflects the ability of RPM3 to flexibly assign the tasks to the workers based on their available resources. In addition, this property reflects the fact that RPM3 can design tasks to have arbitrarily small size to fit the computational power of the available workers. 
 
{\em Comparison to the improved scheme of \cite{kakar2018rate}: }This scheme has a better rate than naively using Lagrange polynomials to send the tasks to the workers. The better rate is achieved by aligning the coefficients in $\ftr{t}{u}(x)$ and $\gtr{t}{u}(x)$ to reduce the number of needed evaluations\footnote{Note that one could use the polynomials of the improved scheme in \cite{kakar2018rate} instead of Lagrange polynomials to improve the rate of the RPM3 scheme. However, the polynomials in \cite{kakar2018rate} require a large number of workers (per cluster in our case) and do not exist for all values of $m$ and $k$.} from the workers. We assume that master sends several tasks to the workers. Each task is of size $m_I k_I$ where $(m_I+z)(k_I+1) - 1 = n-s$ to tolerate $s$ stragglers. The master must send $\lceil mk / m_I k_I\rceil$ tasks to the workers. The rate of this scheme is given by
\begin{equation*}
    \rho_{I} = \left \lceil \dfrac{mk}{m_I k_I} \right\rceil \dfrac{m_I k_I}{(m_I+z)(k_I+1) - 1 }.
\end{equation*}

To compare the rates of RPM3 and the naive scheme we assume that $mk$ divides $m_Ik_I$ and compute the ratio $\rho_I/\rho$
\begin{equation}\label{eq:ratio1}
    \dfrac{\rho_{I}}{\rho} =
     \dfrac{2mk(1+\varepsilon)+ (z-1)\tau_c\sum_{u=1}^c\gamma_{u,c} + z\tau_c \gamma_{1,c}} {(m_I+z)(k_I+1) - 1}.
\end{equation}
Let $D\triangleq m_I k_I - mk(1+\varepsilon)$ be the difference between the number of multiplications needed by the master to obtain $\mA\mB$ when using the improved scheme and RPM3. From~\eqref{eq:ratio1} we deduce that the rate of RPM3 is smaller than the one of \cite{kakar2018rate} when the following holds.

\begin{align*}
     D + m_I+z k_I+z - 1 & \leq mk(1+\varepsilon)+  z\tau_c \gamma_{1,c}
     \nonumber  \\
     &+ (z-1)\tau_c\sum_{u=1}^c\gamma_{u,c}.
\end{align*}
The right hand side of the previous equation is always positive. Therefore, for values of $m_I,k_I,m,k$ and $z$ such that\footnote{In general, we expect $D$ to be negative because RPM3 can generate smaller tasks, i.e., $m\geq m_I$ and $k\geq k_I$.} $m_I+z k_I+z - 1\leq -D$, the improved scheme has a smaller rate than RPM3 independently of the number of clusters and the number of tasks sent to each cluster when using RPM3. However, when $m_I+z k_I+z - 1> -D$, the loss in rate of RPM3 depends on the number of clusters and on the number of tasks sent to each cluster. More precisely, the loss of rate in RPM3 mainly happens due to sending several polynomials per round. 


However, the crucial advantage of RPM3 is the reduced time spent at the master to finish its computation. In RPM3, the master waits until each worker of the slowest cluster computes $\tau_c$ tasks. Whereas, in the scheme of \cite{kakar2018rate} the master waits until every non-straggling worker computes $\lceil mk/ m_I k_I \rceil$ tasks. In particular, assume that the slowest non-straggler in the improved scheme belongs to the slowest cluster in RPM3. If $\tau_c<\lceil mk/m_I k_I \rceil$, then in RPM3 the master waits for the slowest workers to compute a smaller number of tasks which increases the speed of the computation with high probability.
 
 \begin{IEEEproof}[Proof of Lemma~\ref{lemma:rate}] To prove Lemma~\ref{lemma:rate}, we count the number of results $N$ collected by the master at the end of the computation process. From each cluster of workers $u$, $u=1,\dots,c$ the master collects $n_u \tau_u$ results. Recall that $n_1= 2d_1 +2z -1$ and $n_u = 2d_u+z-1$ for $u=2,\dots,c$. We can write the following
 \begin{align}
     N & = \sum_{u=1}^{c} n_u \tau_u \nonumber\\
     & = \sum_{u=2}^{c} (2d_u + z -1)\tau_u + (2d_1 + 2z -1)\tau_1 \nonumber\\
     & = \sum_{u=1}^{c} 2d_u\tau_u + (z-1)\sum_{u=1}^{c} \tau_u + z\tau_1 \nonumber\\
     & = 2mk(1+\varepsilon)+ (z-1)\tau_c\sum_{u=1}^c\gamma_{u,c} + z\tau_c \gamma_{1,c}. \label{eq:sumoftasks}
 \end{align}

 Equation~\eqref{eq:sumoftasks} follows from the fact that $\sum_{u=1}^{c} d_u\tau_u = mk(1+\varepsilon)$. This is true because the master needs $mk(1+\varepsilon)$ different values of $\fcA{i}{t}{u}\fcB{j}{t}{u}$ in total to compute $\mA\mB$ and each interpolated polynomial $\htr{t}{u}(x)$ encodes $d_u$ such values.
 \end{IEEEproof}



\section{Conclusion}\label{sec:conc}
We consider the heterogeneous setting of the private distributed matrix-matrix multiplication. The workers have different resources that are time-varying. We design a scheme called RPM3 that allows the master to group the workers in clusters of workers with similar resources. Each cluster of workers is assigned a number of tasks proportional to the resources available to the workers, i.e., faster workers compute more tasks and slower workers compute less tasks. This flexibility increases the speed of the computation.

In the special case where the resources of the workers in different clusters are proportional to each other, we show that there exists a tradeoff between the flexibility of the RPM3 and its rate. We believe that this tradeoff holds true in the general setting as well. 
We leave the time analysis of the RPM3 scheme as an interesting open problem. Such analysis provides a better understanding the effect of reducing the rate (increasing the number of required tasks from the workers) on the speed of the computation.

\bibliographystyle{ieeetr}
\bibliography{IEEEabrv,PDMM}

\appendix
We want to prove that every round $t$, the tasks sent to the workers do not reveal any information about the input matrices $\mA$ and $\mB$. Recall that we define $\cW_{i,t}$ as the set of random variables representing the tasks sent to worker $w_i$ at round $t$. In addition, for a set $\mathcal{A}\subseteq [n]$ we define $\mathcal{W}_{\mathcal{A},t}$ as the set of random variables representing the tasks sent to the workers indexed by $\mathcal{A}$ at round $t$, i.e., $\mathcal{W}_{\mathcal{A},t}\triangleq \{\cW_{i,t}| i\in \mathcal{A}\}$. The privacy constraint is then expressed as
\begin{equation*}
    {I}\left(\rvA,\rvB;\cW_{\mathcal{Z},t}\right) = 0, \forall \cZ \subset [n], \text{ s.t. } |\cZ| = z.
\end{equation*}

We start by proving the privacy constraint for $\mA$. For a set $\cA\subseteq[n]$, let $\cF_{\cA,t}$ be the set of random variables representing the evaluations of $\ftr{t}{u}(x)$ sent to workers indexed by the set $\cA$ at round $t$. We want to prove 
\begin{equation*}
    {I}\left(\rvA;\cF_{\mathcal{Z},t}\right) = 0, \forall \cZ \subset [n], \text{ s.t. } |\cZ| = z.
\end{equation*}
Proving the satisfaction of the privacy constraint for $\mB$ follows the same steps and is omitted.

Let $\cK$ be the set of random variable presenting the random matrices $\mR_{t,1},\dots,\mR_{t,z}$ generated by the master at round $t$. We start by showing that proving the privacy constraint is equivalent to proving that $H(\cK \mid \cF_{\cZ}, \rvA) = 0$ for all $\cZ\subseteq [n], |\cZ| = z$. The explanation of $H(\cK \mid \cF_{\cZ}, \rvA) = 0$ is that given the matrix $A$ and all the tasks received at round $t$, any collection of $z$ workers can obtain the value of the matrices $\mR_{t,1},\dots,\mR_{t,z}$. To that end we write,
\begin{align}
H(\rvA \mid \cF_{\cZ})&=H(\rvA)-H(\cF_{\cZ})+H(\cF_{\cZ} \mid \rvA)\\
~&=H(\rvA)-H(\cF_{\cZ})+H(\cF_{\cZ} \mid \rvA) \nonumber\\
& \qquad ~~~  -H(\cF_{\cZ} \mid \rvA, \cK) \label{eq:sk}\\
~&=H(\rvA)-H(\cF_{\cZ})+I(\cF_{\cZ}; \cK \mid \rvA) \nonumber \\
~&=H(\rvA)-H(\cF_{\cZ})+H(\cK \mid \rvA) - H(\cK \mid \cF_{\cZ},\rvA) \nonumber\\
~&=H(\rvA)-H(\cF_{\cZ})+H(\cK) - H(\cK \mid \cF_{\cZ},\rvA) \nonumber \\
~&=H(\rvA) - H(\cK \mid \cF_{\cZ},\rvA). \label{eq:keys}
\end{align}
\vspace{0.2cm}

\noindent Equation~\eqref{eq:sk} follows because $H(\cF_{\cZ} \mid A, \cK)=0$, i.e., the tasks sent to the workers are a function of the matrix $\mA$ and the random matrices $\mR_{t,1},\dots,\mR_{t,z}$ which is true by construction. In~\eqref{eq:keys} we use the fact that the random matrices $\mR_{t,1},\dots,\mR_{t,z}$ are chosen independently from $\mA$, i.e., $H(\cK \mid \rvA)=H(\cK)$. Equation~\eqref{eq:keys} follows because for any collection of $z$ workers, the master assigns $z$ tasks each of which has the same dimension as $\mR_{t,\delta}$, $\delta\in \{1,\dots,z\}$. In addition, all matrices $\mR_{t,1},\dots,\mR_{t,z}$ are chosen independently and uniformly at random; hence, $H(\cF_{\cZ}) = H(\cK)$. 

Therefore, since the entropy $H(.)$ is positive, proving that $H(\rvA\mid \cF_{\cZ}) = H(\rvA)$ is equivalent to proving that $H(\cK \mid \cF_{\cZ},A) = 0$.

The remaining part of the proof is to show that given the matrix $A$ and all the tasks received at round $t$, any collection of $z$ workers can obtain the value of the matrices $\mR_{t,1},\dots,\mR_{t,z}$. This follows immediately from the use of Lagrange polynomials and setting the random matrices as the first $z$ coefficients. More precisely, given the data matrix as side information, the tasks sent to any collection of $z$ workers become the evaluations of a Lagrange polynomial of degree $z-1$ whose coefficients are the random matrices $\mR_{t,1},\dots,\mR_{t,z}$. Thus, the workers can interpolate that polynomial and obtain the random matrices.
\end{document}